\documentclass[prl,twocolumn]{revtex4}
\usepackage{graphicx}
\usepackage{amssymb}
\usepackage{color}
\usepackage{enumerate}
\usepackage{amsmath}
\usepackage{graphics}
\newcommand{\be}{\begin{equation}}
\newcommand{\ee}{\end{equation}}
\newcommand{\bea}{\begin{eqnarray}}
\newcommand{\eea}{\end{eqnarray}}

\begin{document}

\title{Collective behavior of light in vacuum}

\author{Fabio Briscese}\email{briscese.phys@gmail.com, briscesef@sustc.edu.cn}

\affiliation{Department of Physics, Southern University of Science
and Technology, Shenzhen 518055, China\\}

\affiliation{ Istituto Nazionale di Alta Matematica Francesco
Severi, Gruppo Nazionale di Fisica Matematica, Citt\`{a}
Universitaria, P.le A. Moro 5, 00185 Rome, Italy.}

\begin{abstract}

Under the action of light-by-light scattering, light beams show
collective behaviors  in vacuum. For instance, in the case of two
counterpropagating laser beams with specific initial helicity, the
polarization of each beam oscillates periodically between the left
and right helicity. Furthermore, the amplitudes and the
corresponding intensities of each polarization propagate like
waves. Such polarization waves might be observationally accessible
in future laser experiments, in a physical regime complementary to
those explored by particle accelerators.

\end{abstract}

\maketitle

\section{Introduction}\label{introduction}

Light self-interaction is a purely quantum effect, since the
classical Maxwell equations are linear, and this forbids processes
such as light-by-light scattering ($\gamma \gamma \rightarrow
\gamma \gamma$) that are allowed in quantum electrodynamics.
Indirect evidence of such processes has been found in particle
accelerators
\cite{denterria1,denterria2,denterria3,denterria4,denterria5,denterria6,denterria7,atlas},
while the search for signatures of light-by-light scattering in
optics is still in progress
\cite{lammerzal,jose,pike,Dinu:2014tsa,Dinu:2013gaa,PVLAS,cadene,winstisen,king,di
piazza1,di piazza2,di piazza3,di piazza4,di piazza5,di piazza6,di
piazza7,di piazza8,di piazza9,di piazza10,di piazza11,di
piazza12,di piazza13,di piazza14,di piazza15,di piazza16,di
piazza17,di piazza18,di piazza19,di piazza20,di piazza21,di
piazza22,di piazza23,di piazza24,di piazza 25,di piazza26,di
piazza27,di piazza28,di piazza29,di piazza30,di piazza31,di
piazza32,king new,Briscese PLA}. However, this situation might be
overcome in near future. In facts, it has been shown
\cite{Briscese PLA} that, despite their weakness, quantum
corrections due to light-by-light scattering can change
dramatically the dynamics of the electromagnetic field, inducing
effects that can be tested experimentally.


Quantum corrections to Maxwell equations  have been calculated a
long time ago by Heisenberg and Euler \cite{euler}, and
extensively studied by other authors
\cite{dicus,Karplust,leo,schwinger}. The effective Lagrangian of
the electromagnetic field, obtained retaining only the dominant
one electron loop corrections \footnote{We are neglecting other
contributions to light-by-light scattering, such as those due to
the $\mu$ and $\tau$ loops, which are suppressed by a factor $\sim
\left(m_e/m_\mu\right)^4 \, \epsilon^2 \, F_{\mu\nu} F^{\mu\nu}$
and $\sim \left(m_e/m_\tau\right)^4 \, \epsilon^2 \, F_{\mu\nu}
F^{\mu\nu}$ respectively.}, is \cite{schwinger}

\begin{equation}\label{lagrangian}
L = \frac{1}{4} F_{\mu\nu}F^{\mu\nu} + \epsilon^2 \left[ \left(
F_{\mu\nu}F^{\mu\nu}\right)^2 - \frac{7}{16} \left(
F_{\mu\nu}\tilde F^{\mu\nu}\right)^2 \right] \, ,
\end{equation}
where $F^{\mu\nu}= A^{\mu,\nu}-A^{\nu,\mu}$ is the electromagnetic
field \footnote{We use the covariant formalism, so that the zeroth
coordinate is defined as $x^0 = c \, t$.}, $A^\mu$ is the
electromagnetic four-potential, $\tilde F^{\mu\nu} \equiv
\epsilon^{\mu\nu\alpha\beta}F_{\alpha\beta}$,  $\epsilon^2 =
\alpha^2 \left(\hbar/m_e c \right)^3/90 m_e c^2$,   $\alpha = e^2
/4\pi \epsilon_0 \hbar c\simeq 1/137$ is the fine structure
constant, $\epsilon_0$  the dielectric permeability of vacuum,
$m_e$ the electron mass and $c$ the speed of light.

In this paper we analyze the effect of light-by-light scattering
on the dynamics of the electromagnetic field in vacuum. Hereafter,
we consider low energetic photons with energies $ \ll m_e c^2$, so
that particles creation is inhibited, and light-by-light
scattering is the only process involving photons. Under these
hypothesis, the Lagrangian (\ref{lagrangian}) is fit for our
purpose.

The terms $\sim \epsilon^2$ in (\ref{lagrangian}) account for
light-by-light scattering, introducing cubic corrections in the
equations for the four-potential $A^\mu$. Since $\epsilon^2 \sim 4
\times 10^{-31} m^3/J$, one has $\epsilon^2 \,
F_{\mu\nu}F^{\mu\nu} \ll 1$ and $\epsilon^2 \,F_{\mu\nu}\tilde
F^{\mu\nu} \ll 1 $ in realistic physical conditions, so that
nonlinear corrections to Maxwell equations are usually negligible
\footnote{We note that, the next-to-leading terms in the expansion
of the Heisenberg-Euler Lagrangian are suppressed by a factor
$\sim \left( \epsilon^2 \, F_{\mu\nu} F^{\mu\nu}\right)^2$; indeed
they are negligible to any extent in realistic laboratory
conditions for low energetic ($E \ll m_e \, c^2$) photons. What is
more, they are subdominant with respect to contributions due to
other channels in light-by-light scattering, e. g., the $\mu$ and
$\tau$ loops.}. However, this is not the case for some specific
configurations of the electromagnetic field, that become unstable
due to the action of hidden resonances. In fact, in \cite{Briscese
PLA} it has been shown that such tiny nonlinearities affect
heavily the polarization of the electromagnetic waves in vacuum;
indeed their polarization oscillates periodically in time between
right and left helicity states.

Here we extend this result, which has been obtained for
counterpropagating homogeneous (in space) plane waves, to more
general configurations. Such extension is mathematically
straightforward, but its physical implications are relevant. We
show that the polarization oscillations occur both in space and
time. It is found that the amplitudes of the different
polarizations, and  the corresponding intensities, propagate as
plane waves.  The occurrence of super-luminal polarization waves
is considered, and possible contradictions with special
relativity, and their solution, are discussed. Finally, we discuss
the possibility of observing polarization waves in laser
experiments and argue how the recurrence time of the polarization
oscillations can be reduced, in order to favor their detection.

The importance of these results is in the fact that they show that
light exhibit  collective behaviors  in vacuum, which are
triggered by light-by-light scattering. Such collective modes are
represented by polarization waves, whose properties  have been
completely characterized. Remarkably, this phenomenology has been
obtained analytically through a simple multiscale approach, as
described below.

\section{Multiscale equations}

Starting from the Lagrangian (\ref{lagrangian}) it is easy to show
that the modified Maxwell equations for the electromagnetic
four-potential $A^\alpha$ in the Lorentz gauge ($\partial_\alpha
A^\alpha = 0$) are

\begin{equation}\label{maxwell eq}
\begin{array}{ll}
\Box A^\alpha \left(1 + 8 \, \epsilon^2 \,
F_{\mu\nu}F^{\mu\nu}\right) + \epsilon^2 \, B^\alpha = 0 \, ,
\end{array}
\end{equation}
where $\Box$ is the d'Alembertian operator and
\begin{equation}\label{definition B}
\begin{array}{ll}
B^\alpha \equiv 8 \,  \left[  F^{\alpha\beta}
\partial_\beta \left(F_{\mu\nu}F^{\mu\nu}\right) -  \frac{7}{16}
\tilde F^{\alpha\beta}
\partial_\beta \left( F_{\mu\nu}\tilde F^{\mu\nu}\right) \right] \, .
\end{array}
\end{equation}

At zeroth order in $\epsilon$  Eq. (\ref{maxwell eq}) reduces to
the Maxwell equations in vacuum $\Box A^{\alpha} = 0$, which can
be solved exactly. Let us consider a zeroth order solution
$A^{(0)\alpha}$ corresponding to a system of two plane
electromagnetic waves propagating in the $x^3$ direction. With a
proper gauge choice we set $A^{(0)0} = A^{(0)3} = 0$, so that we
can write the four potential in a more convenient vector form as

\begin{equation}\label{plane waves order 0}
\vec{A}^{(0)} = \vec{a} + \vec{b} + c.c. \, ,\\
\end{equation}
(where c.c. stands for complex conjugate), i.e., as the
superposition of the two plane waves $a$ and $b$ defined as
\begin{equation}\label{plane waves order 0 polarizations}
\begin{array}{ll}
\vec{a} =  \left( a_L \, \hat e_L + a_R \, \hat e_R \right)  \,
e^{i k x} \, ,  \vec{b} =  \left(b_L \, \hat e_L + b_R \, \hat e_R
\right) e^{i h x} \, ,
\end{array}
\end{equation}
with the two  wave vectors  $k=(k_0,0,0,k_3)$ and
$h=(h_0,0,0,h_3)$ satisfying the dispersion relation $|k_0/k_3|=
|h_0/h_3|=1$. Here $\hat e_L = (1,i,0)/\sqrt{2}$ and $\hat e_R =
(1,-i,0)/\sqrt{2}$ are the left and right polarization vectors.
Therefore, the coefficients $a_L$, $a_R$, $b_L$ and $b_R$ are the
complex amplitudes of the left and right polarizations of the
plane waves $a$ and $b$, and their squared modules are
proportional to the corresponding intensities.

Let us study how the dynamics or the plane waves (\ref{plane waves
order 0}) are modified due to the effect of quantum corrections.
We anticipate that such dynamics  entails  slow variations of the
complex amplitudes in space and time. The smallness of
$\epsilon^2$ suggests that a complete solution of (\ref{maxwell
eq}) might be obtained through a standard perturbative expansion
of the four-vector in powers of $\epsilon^2$. However, such a
naive approach fails  when the two waves $a$ and $b$ are
counterpropagating, e.g. when $k_0/k_3= -h_0/h_3=1$, due to the
occurrence of secular divergences of small perturbations. In fact,
in such a configuration, evaluating $B^\alpha$ over the solution
(\ref{plane waves order 0}) one has \cite{Briscese PLA}
\begin{equation}\label{second order B}
\begin{array}{ll}
\vec{B}  = 32 \, k_0^2 \, h_0^2 \times \\
\\
\left\{ \left[ \left( -3 \, a_L \left(|b_L|^2+|b_R|^2 \right) + 22
\, a_R
b_L \bar b_R \right) \hat e_L + \right.\right.\\
\\
\left.\left. \left( -3 \, a_R \left(|b_L|^2+|b_R|^2 \right) + 22
\,
a_L b_R \bar b_L \right) \hat e_R \right] e^{i k x} \right.\\
\\
\left[ \left( -3 \, b_L \left(|a_L|^2+|a_R|^2 \right) + 22 \, b_R
a_L \bar a_R \right) \hat e_L + \right.\\
\\
\left.\left. \left( -3 \, b_R \left(|a_L|^2+|a_R|^2 \right) + 22
\,  b_L a_R \bar a_L \right) \hat e_R \right] e^{i h x} \right\}+\\
\\
+ \, c. \, c. \, + \, $nonresonant terms$ \, .
\end{array}
\end{equation}
Therefore $\vec{B}$ contains  resonant terms $\sim e^{i k x}$ and
$\sim e^{i h x}$ that make any small perturbation of (\ref{plane
waves order 0}) diverge as $\delta A \sim \epsilon^2 \, t \, c \,
k^3 (A^{(0)})^3 \, e^{i k x}$,  where we have assumed for
simplicity that $k_0 \sim h_0\sim k$. See \cite{Briscese PLA} for
a discussion of the secular divergence of small perturbations of
(\ref{plane waves order 0}) in this configuration.

The emergence of secularities and the consequent failure of
perturbative power expansions is quite common in physics. Usually,
this happens in problems in which the solutions depend
simultaneously on widely different scales. In such cases, the
divergences can be handled through  a multiscale expansion,
introducing suitable slow variables; see \cite{multiscale book}
for an introduction to the multiscale perturbative method. As we
will see, the multiscale approach provides an approximate solution
of (\ref{maxwell eq}) that captures all the essential features of
the problem under analysis.

We introduce the slow variable $y^0$ as

\begin{equation}\label{multiscale variables}
y^0 \equiv \epsilon^2 \left(n_0 \, x^0 + n_3 \, x^3 \right) \, ,
\end{equation}
where $n_0,n_3 \in\mathbb{R}$ are the covariant components of a
four-vector $n = \left(n^0,0,0,n^3\right)$, so that the
relativistic covariance of (\ref{multiscale variables}), as well
as that of the multiscale approximate solutions, is preserved. We
choose $n$ dimensionless, so that $y^0$ is measured in $m^4/J$.
The multiscale treatment requires that $n_0$ and $n_3$ are such
that $|n_0| + |n_3| \sim 1$. Moreover, to have meaningful
multiscale equations, it will be necessary to impose the condition
$n_0 \neq \pm n_3$. Using (\ref{multiscale variables}) one has
$\partial_{x^0} \rightarrow
\partial_{x^0} +  \epsilon^2 \, n_0 \, \partial_{y^0}$ and
$\partial_{x^3} \rightarrow \partial_{x^3} +  \epsilon^2 \, n_3 \,
\partial_{y^0}$, which finally gives the d'Alembertian in terms of
the derivatives with respect to slow and fast variables as

\begin{equation}\label{multiscale d alembertian}
\begin{array}{ll}
\Box \rightarrow \Box + 2 \epsilon^2 \left( n_0 \, \partial_{x^0}
- n_3 \, \partial_{x^3}\right) \partial_{y^0} +
o\left(\epsilon^4\right)\, .
\end{array}
\end{equation}
We split the dependence of the four potential into slow and fast
variables, assuming that the amplitudes $a_L$, $a_R$, $b_L$ and
$b_R$  depend only on the slow variable $y^0$. We search the
solutions of the Eq.s (\ref{maxwell eq}) in the form $\vec A =
\vec A^{(0)} + \epsilon^2 \delta \vec A$, so that at order $\sim
\epsilon^2$ Eq. (\ref{maxwell eq}) gives

\begin{equation}\label{multiscale equations}
\begin{array}{ll}
2 \left( n_0 \, \partial_{x^0} - n_3 \,
\partial_{x^3}\right) \partial_{y^0} \vec A^{(0)} +
\Box \delta \vec A + \vec B = 0
\end{array}
\end{equation}
The multiscale equations are obtained by imposing that the first
term in (\ref{multiscale equations}) cancels the resonant terms in
$\vec B$, while $\Box \delta \vec A$ equals the remaining non
resonant terms, so that the small perturbation $\delta \vec A$ is
stable. In that way, we obtain the dynamical equations for the
complex amplitudes as

\begin{equation}\label{multiscale equations amplitudes}
\begin{array}{ll}
i  a^\prime_L + 16 \frac{k_0 h_0^2}{n_0-n_3} \left( -3 \, a_L
\left(|b_L|^2+|b_R|^2 \right) + 22 \, a_R
b_L \bar b_R \right)   = 0\\
\\
i  a^\prime_R + 16 \frac{k_0 h_0^2}{n_0-n_3} \left( -3 \, a_R
\left(|b_L|^2+|b_R|^2 \right) + 22 \, a_L b_R \bar b_L \right)   =
0\\
\\
i  b^\prime_L + 16 \frac{k_0^2 h_0}{n_0+n_3} \left( -3 \, b_L
\left(|a_L|^2+|a_R|^2 \right) + 22 \, b_R
a_L \bar a_R \right)   = 0\\
\\
i   b^\prime_R + 16 \frac{k_0^2 h_0}{n_0+n_3} \left( -3 \, b_R
\left(|a_L|^2+|a_R|^2 \right) + 22 \, b_L a_R \bar a_L \right) = 0
\, ,
\end{array}
\end{equation}
where $f^\prime\equiv df/dy^0$. It is now evident why the
condition $n_0 \neq \pm n_3$ is necessary in order to have
meaningful multiscale equations.

\section{Results}

Let us study (\ref{multiscale equations amplitudes}) in detail.
First of all, it is quite immediate to recognize that the energy
densities $<\rho_a> = k_0^2 \left(|a_L|^2+|a_R|^2 \right)$ and
$<\rho_b> = h_0^2 \left(|b_L|^2+|b_R|^2 \right)$ of the two beams
$a$ and $b$ are constant; therefore, the intensities of the two
plane waves $a$ and $b$ are conserved separately. Furthermore, the
quantity $S = k_0 \left(n_0-n_3 \right) \left(|a_L|^2-|a_R|^2
\right)+ h_0 \left(n_0-n_3 \right)\left(|b_L|^2-|b_R|^2 \right)$,
that in the case $n_0=1$ and $n_3=0$ corresponds to the spin
density, is also conserved \footnote{It is worth mentioning that
$<\rho_a>$, $<\rho_b>$ and $S$ are the  zeroth order
approximations of the energy and spin densities in the nonlinear
classical theory \cite{di piazza27}, and they coincide with the
corresponding quantities in perturbative quantum field theory.
This is reasonable, since Eq.s (\ref{multiscale equations
amplitudes}) have been obtained in perturbation theory, and the
quantum corrections have been calculated in perturbative quantum
field theory.}. Exploiting these relations, the system
(\ref{multiscale equations amplitudes}) can be simplified and then
resolved exactly \cite{Briscese PLA}, showing that the evolution
of the modules of the complex amplitudes is periodic.

We can now estimate the period $\Delta y^0$  of the polarization
oscillations. This is roughly given by the scale  at which the
amplitudes changes significantly. Assuming that $a_L \sim a_R \sim
a_0$ and $b_L \sim b_R \sim b_0$, such a scale is given by the
conditions $\Delta a_{L,R}/a_{L,R} \sim 1$ and $\Delta
b_{L,R}/b_{L,R} \sim 1$, which, using (\ref{multiscale equations
amplitudes}), gives the two scales $\Delta y^0_a \sim
\frac{|n_0+n_3|}{k_0 h_0^2 |b_0|^2}$ and $\Delta y^0_a \sim
\frac{|n_0-n_3|}{k_0^2 h_0 |a_0|^2}$. Since the solutions of
(\ref{multiscale equations amplitudes}) are periodic in $y^0$,
$\Delta y^0$ will be  the minimum between $\Delta y^0_a$ and
$\Delta y^0_b$, i.e.

\begin{equation}\label{period y0}
\Delta y^0 \sim inf \left\{\frac{|n_0+n_3|}{k_0 h_0^2 |b_0|^2},
\frac{|n_0-n_3|}{k_0^2 h_0 |a_0|^2} \right\} \, .
\end{equation}

At this point we can characterize the dynamics of the system. Some
of the solutions of (\ref{multiscale equations amplitudes}) are
easily found. In fact, if $a_L \, a_R = 0$ and $b_L \, b_R = 0$ at
$y^0 =0$ so that the waves $a$ and $b$ have circular polarization,
such products remain always zero and the solutions of
(\ref{multiscale equations amplitudes}) are  complex exponentials
$\sim e^{i \omega y^0}$. This is true also in the case of two
linearly polarized waves with $a_L = \pm a_R$ and $b_L = \pm b_R$.
However, such solutions are not interesting, since the modules of
the complex amplitudes remains constant, and the effect of
light-by-light scattering is just a negligibly small $\sim
\epsilon^2$ correction to the dispersion relation of the plane
waves $a$ and $b$ (See \cite{Briscese PLA} for possible
implications for quantum gravity phenomenology \cite{amelino}).

Other solutions show a behavior much more interesting, since the
polarizations of the light beams change dramatically during the
evolution of the system. These solutions correspond to initial
conditions such that at least one of the products $a_L \, a_R$ or
$b_L \, b_R$ is  different from zero. This is due to the fact
that, when nonzero, the last terms in Eq.s (\ref{multiscale
equations amplitudes}) are responsible for the oscillatory
behavior that we describe below. The system (\ref{multiscale
equations amplitudes}) can be solved analytically \cite{Briscese
PLA}; however, for our purposes it will be sufficient to discuss
numerical solutions. Solving (\ref{multiscale equations
amplitudes}) numerically, it is possible to see that the
polarizations of the two counterpropagating waves oscillate
periodically between left and right  configurations.

\begin{figure}[tbp]
\begin{center}
\includegraphics[width=3.4in, scale=1]{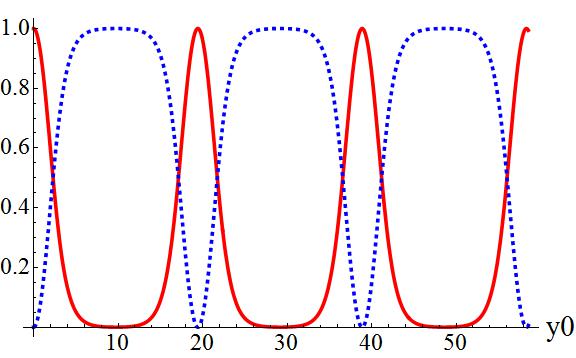} \caption{ We plot the
evolution  of  $|a_L|^2/|a^0_L|^2+|a^0_R|^2$ (solid  red line) and
$|a_R|^2/|a^0_L|^2+|a^0_R|^2$ (dashed blue line) against $y^0$ (in
units of $m^4/J$). The plot shows the oscillatory behavior of the
polarization of the light beam $a$. } \label{fig1}
\end{center}
\end{figure}

\begin{figure}[tbp]
\begin{center}
\includegraphics[width=3.4in, scale=1]{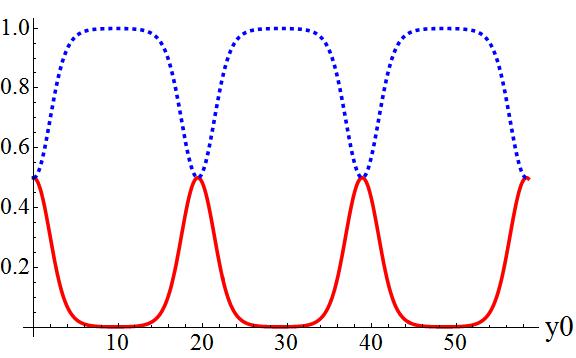} \caption{ We plot the
evolution  of  $|b_L|^2/(|b^0_L|^2+|b^0_R|^2)$ (solid red line)
and $|b_R|^2/(|b^0_L|^2+|b^0_R|^2)$ (dashed blue line) against
$y^0$ (in units of $m^4/J$). The plot shows the oscillatory
behavior of the polarization of the light beam $b$. } \label{fig2}
\end{center}
\end{figure}

This effect is particularly evident when only one of the waves $a$
and $b$ is initially polarized circularly, e.g. $a_L \, a_R = 0$
and $b_L \, b_R \neq 0$. For instance, we solve (\ref{multiscale
equations amplitudes}) for $k_0=h_0= 0.1$ and initial values
$a^0_R = 0$, $a^0_L = 1$, $b^0_L = 1$, $b^0_R = i$, $n_0 =
10^{-3}$ and $n_3=1$. From Fig. \ref{fig1} we see that $|a_R|$ is
initially zero but it grows   to $|a_R|=|a^0_L|$, while $|a_L|$
goes  to zero. Thus, the beam $a$ is initially in the left-handed
polarization, but then it    switches to the right-handed
polarization. It remains in this state  until it jumps back to its
initial left-handed configuration after the first period. The
behavior of the beam $b$ is similar. In fact, $|b_L|$ goes to
zero, while $|b_R|$ goes to $\sqrt{|b^0_L|^2+|b^0_R|^2}$, and
after the first period $|b_L|$ and $|b_R|$  go back to their
initial values. The difference with respect the beam $a$ is that
$|b_L|$ never reaches the zero. From Fig.s \ref{fig1} and
\ref{fig2} it is also evident that the evolution of the modules of
the complex amplitudes is periodic.

Numerical investigation of (\ref{multiscale equations amplitudes})
shows that the oscillatory behavior of the system is not affected
(qualitatively) by the choice of the parameters in
(\ref{multiscale equations amplitudes}), while the period of the
oscillations depends on such parameters (in order magnitude) as in
(\ref{period y0}). For instance, for the solution plotted  in
Fig.s \ref{fig1} and \ref{fig2}, Eq. (\ref{period y0}) gives a
period $\Delta y^0 \sim 3.9$ which is a good estimation (as an
order of magnitude) of the actual period  $\simeq 20$, as seen in
the plots.

At that point, it becomes necessary to discuss the physical
meaning of the two parameters $n_0$ and $n_3$. Such parameters are
not fixed by the multiscale, except for the conditions $|n_0| +
|n_3| \sim 1$ and $n_0 \neq \pm n_3$. The freedom in their choice
reflects the fact that  our multiscale solution is not the general
solution of (\ref{maxwell eq}), but it still contains some
residual freedom in the choice of the initial values of the
derivatives of the complex amplitudes of $a$ and $b$.

For instance, imposing $n_0 = 1$ and $n_3 =0$, we have $y^0 =
\epsilon^2 \, x^0$, so that the amplitudes are homogeneous in
space and periodic in the slow variable $y^0 = \epsilon^2 \, c \,
t$, indeed periodic in time. This class of solutions has been
discussed extensively in \cite{Briscese PLA}. On the contrary, the
choice $n_0 = 0$ and $n_3 =1$ corresponds to  static solutions
that are periodic in space, since in this case $y^0 = \epsilon^2
\, x^3$.

In general, the complex amplitudes are periodic functions of the
slow variable $y^0 = \epsilon^2 \left(n_0 \, x^0 + n_3 \, x^3
\right)$; therefore they propagate as plane waves with speed $v_p
= |n_0/n_3|$. A first remark is that the superposition principle
is not valid for these waves, since the system (\ref{multiscale
equations amplitudes}) is nonlinear.  Furthermore, it must be
emphasized that that such "polarization waves" can not travel at
the speed of light, since it must be $n_0 \neq \pm n_3$, while
they can -- at least in principle -- travel faster than light when
$ |n_0/n_3|>1$.

It is not evident that the existence of super-luminal polarization
waves is in contradiction with special relativity. For instance,
there is no manner to control the polarization waveform, and
therefore encode information that can travel faster than light.
Instead,  light beams self-organize in such a way that their
polarizations evolve as waves, which might propagate faster than
light. What is more, since the partial intensities of the two
beams are conserved separately, polarization waves do not carry
energy. Moreover, it is know that the group velocity of a light
beam, i.e., the velocity of its envelope, exceeds the speed of
light in some circumstances \cite{group velocity}; but also in
this case there is no contradiction with special relativity, since
there is no propagation of signals or energy with a velocity above
$c$.

However, it might result that the existence of super-luminal
polarization waves contradicts special relativity.  In such
eventuality, these superluminal polarization waves must be
considered unphysical, and we must impose the condition $|n_0/n_3|
< 0$; the meaning of these conditions would be that we should
avoid unphysical initial conditions.

We mention that the search for the effects of light
self-interactions in optics is already under study
\cite{lammerzal,jose,pike,Dinu:2014tsa,Dinu:2013gaa,PVLAS,cadene,winstisen,king,di
piazza1,di piazza2,di piazza3,di piazza4,di piazza5,di piazza6,di
piazza7,di piazza8,di piazza9,di piazza10,di piazza11,di
piazza12,di piazza13,di piazza14,di piazza15,di piazza16,di
piazza17,di piazza18,di piazza19,di piazza20,di piazza21,di
piazza22,di piazza23,di piazza24,di piazza 25,di piazza26,di
piazza27,di piazza28,di piazza29,di piazza30,di piazza31,di
piazza32,king new,Briscese PLA}.  A review of solutions in
nonlinear QED is given in \cite{di piazza20}. Moreover, the
self-interactions of magnetic and electric moments have been
studied in \cite{di piazza21}, linear and nonlinear responses of
constant background to electric charge have been studied in
\cite{di piazza22,di piazza23,di piazza24}, the linear response in
the form of magnetic monopole has been studied in \cite{di piazza
25,di piazza26},  and the finiteness of the self-energy of the
point-like charge has been analyzed in \cite{di piazza27,di
piazza28,di piazza29,di piazza30}.

It is also worth  mentioning that the  interaction of two
counterpropagating plane waves under the action of light-by-light
scattering was already studied in \cite{king new}. In that paper
the authors used the standard perturbation theory to study the
evolution of initially small perturbations $\Delta E$ over a
background electromagnetic field $E^{(0)}$. They found that at
some finite time the perturbations $\Delta E$ become dominant over
the background, i. e., they incidentally find the divergence of
perturbations due to secularities discussed in the Introduction
and outlined  in \cite{Briscese PLA}. However, when the (initially
small) perturbations overcome the background,  the perturbative
solution is no longer valid.  Indeed, such solution does not
uncover the oscillatory behavior of polarizations, since this
effect appears only on long time scales, when standard
perturbation theory is unapplicable and one must recur to
multiscale perturbation expansion.

The novelty of the results reported here and in \cite{Briscese
PLA} is that, by means of multiscale analysis, we have obtained
precise analytical results enlightening the most important
features of the collective behavior of light in vacuum induced by
light-by-light scattering. Moreover, we have understood that we
have to look at the polarization rather than at light intensity,
and we know that the case of two counterpropagating laser beams is
the best configuration to observe polarization oscillations.

Let us analyze the observational aspects of the polarization
waves. To have an idea of the observation time required to reveal
the polarization waves, we  estimate their recurrence time $T$ for
light beams produced in petawatt class lasers. The intensities
attainable in these lasers reaches $I \sim 10^{23} W/cm^2$
\cite{lasers1,lasers2}. Thus, according to (\ref{period y0}) the
recurrence time $T \sim \Delta y^0/c$ will be

\begin{equation}\label{period T}
\begin{array}{ll}
T  \sim inf\left\{|n_0+n_3|, |n_0-n_3|\right\}
\left(\epsilon^2 \, k_0 \,  I\right)^{-1} \sim \\
\\
\sim 4 \times 10^2 \times inf\left\{|n_0+n_3|, |n_0-n_3|\right\}
\left(\lambda/m \right) \, s \, ,
\end{array}
\end{equation}
where $\lambda/m$ is the laser wavelength in meters  (we used $k_0
\sim h_0 \sim 2 \pi/\lambda$ and $k_0^2 a^2 \sim k_0^2 b^2 \sim
<\rho> \sim I/c$). Therefore, for $|n_0+n_3| \sim |n_0-n_3| \sim
1$   and $\lambda \sim 1 \, \mu m$ \cite{lasers1,lasers2},
observation time is of the order of $4 \times 10^{-4}\, s$.

\begin{figure}[tbp]
\begin{center}
\includegraphics[width=3.4in, scale=1]{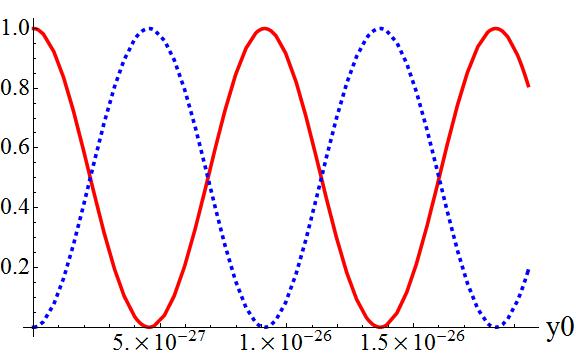} \caption{ We plot the
evolution  of  $|a_L|^2/|a^0_L|^2+|a^0_R|^2$ (solid red line) and
$|a_R|^2/|a^0_L|^2+|a^0_R|^2$ (dashed blue line) against $y^0$ (in
units of $m^4/J$) for $n_0 = 1$, $n_3 = 2$ and $|a^0_L|^2  = 10^3
J/m$, $a^0_R = 0$, $|b^0_L|^2 = |b^0_R|^2 = 10^3 J/m$, $k_0 = h_0
= 10^7 \, m^{-1}$. } \label{fig4}
\end{center}
\end{figure}

This estimation is confirmed numerically. For instance, in Fig.
\ref{fig4} we plot $|a_L|^2/|a^0_L|^2+|a^0_R|^2$  and
$|a_R|^2/|a^0_L|^2+|a^0_R|^2$ for the solution of (\ref{multiscale
equations amplitudes}) with $n_0 = 1$, $n_3 = 2$ and $|a^0_L|^2  =
10^3 J/m$, $a^0_R = 0$, $|b^0_L|^2 = |b^0_R|^2 = 10^3 J/m$, $k_0 =
h_0 = 10^7 \, m^{-1}$, corresponding to $I \simeq 10^{23} W/cm^2$
and $\lambda \sim 1 \, \mu m$. The value of $\Delta y^0 \simeq
10^{-26} m^4/J$ that can be read from the plot  corresponds to a
period $T = \Delta y^0/\epsilon^2 \, c \simeq 10^{-4} \, s$, which
is in good agreement with (\ref{period T}).

\begin{figure}[tbp]
\begin{center}
\includegraphics[width=3.4in, scale=1]{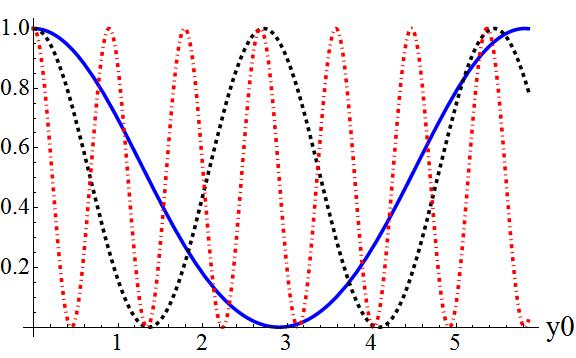} \caption{ We plot the
evolution  of  $|a_L|^2/|a^0_L|^2+|a^0_R|^2$  against $y^0$ (in
units of $m^4/J$) for $\eta = 0.3$ (solid blue line), $\eta = 0.1$
(dashed  black line), $\eta = 0.05$ (dashed-dotted red line) and
$k_0=h_0= 0.1$, $a^0_R = 0$, $a^0_L = 1$, $b^0_L = 1$, $b^0_R =
i$. The plot shows that the period of the oscillations decreases
for decreasing $\eta$. } \label{fig3}
\end{center}
\end{figure}

We stress that the recurrence time can be lowered further,
choosing $n_0$ and $n_3$ in a proper way. In fact, one can make
$T$ smaller while preserving the validity of the multiscale
treatment, e.g. taking $n_0+n_3 = \eta$ and  $n_0-n_3= 1$. From
(\ref{period T}) it is then evident that we can reduce the
recurrence time $T$ choosing $\eta \ll 1$. This fact is confirmed
numerically by solving (\ref{multiscale equations amplitudes}) for
different values of $\eta$. For instance, in Fig. \ref{fig3} we
plot $|a_L|/a_L^0$ as a function of $y^0$ for $\eta = 0.3, \,
0.15, \, 0.05$ and $k_0=h_0= 0.1$, $a^0_R = 0$, $a^0_L = 1$,
$b^0_L = 1$, $b^0_R = i$, showing that the period of the
oscillations decreases for decreasing $\eta$. Indeed,  $T$ is
considerably reduced for $\eta \ll 1$, corresponding to
polarization waves traveling nearly at the speed of light.
However, the practical issue of preparing the system in the proper
initial conditions corresponding to a specific choice of $n_0$ and
$n_3$ remains.

Finally, we  mention that polarization oscillations cannot be
detected in the cosmic microwave background (CMB) radiation
\cite{planck}, since its  energy density $ \sim 10^{-14} J/m^3$
gives extremely small corrections  to the linear dynamics, see
\cite{Briscese PLA}. Moreover, polarization  waves can be of
interest in astrophysics, for instance, they can play a role in
the behavior of magnetized neutron stars \cite{neutron
stars1,neutron stars2,neutron stars3,neutron stars4,neutron
stars5,neutron stars8,neutron stars9}, and in astrophysical
electromagnetic shocks \cite{neutron stars6,neutron stars7};
however we will discuss these issues elsewhere.

\section{Conclusions}

It has been shown that the extremely weak light-by-light
interaction can induce unexpectedly strong deviations from the
free dynamics of light. In particular, it is responsible for the
generation of polarization waves that, in principle, can propagate
faster than light. The phenomenology described above is quite
surprising for different reasons. First, it is a notable example
of how an extremely thin correction, such as that arising from
light-by-light scattering, can produce a strong deviation from the
free dynamics of a system. Furthermore, it is remarkable that
photons have such an ordered behavior, showing a collective
response to the quantum-induced nonlinear effects considered here,
rather than behaving in a chaotic way. Last but not least,
polarization waves might be observationally accessible in laser
experiments, and light-by-light scattering tested in a physical
regime complementary to that explored by particle accelerators.

\textbf{Acknowledgments} The author is very grateful to  F.
Calogero, P. Santini and E. Del Re for useful discussions on the
draft version of this paper.

\end{document}